\documentclass[12pt]{article}
\pdfoutput=1 
\usepackage{slashed}
\usepackage{color, verbatim}
\usepackage{latexsym}
\usepackage{amssymb}
\usepackage{amsmath}
\usepackage{graphicx}
\graphicspath{{figs/}}
\usepackage{arydshln}
\usepackage[dvipsnames,table]{xcolor}
\usepackage{adjustbox}
\usepackage{easybmat}
\usepackage{bbm}
\usepackage{cite}
\usepackage{slashed}
\usepackage{bm}
\usepackage{adjustbox}
\usepackage{booktabs}
\usepackage{multirow} 
\usepackage{rotating}
\usepackage{mathtools}
\usepackage[inline]{enumitem}
\usepackage{lscape}
\usepackage{booktabs}
\usepackage[normalem]{ulem}
\usepackage{subfig}

\usepackage{hyperref}
\definecolor{shadecolor}{rgb}{0.01,0.199,0.1}

\newcommand{\graycell}{\cellcolor[rgb]{0.8,0.8,0.8}}

\newcommand{\ii}{\ensuremath{\mathrm{i}}}

\newcommand{\gtick}{\textcolor{green}{\mathbf{\checkmark}}}
\newcommand{\otick}{\textcolor{orange}{\mathbf{\checkmark}}}
\newcommand{\rtick}{\textcolor{red}{\text{\sffamily X}}}

\newcommand{\mrr}[1]{\textcolor{blue}{#1}}
\newcommand{\mc}[1]{#1}

\renewcommand{\theequation}{\thesection.\arabic{equation}}
\makeatletter
\@addtoreset{equation}{section}
\g@addto@macro\bfseries{\boldmath}
\newcommand\Label[1]{&\refstepcounter{equation}(\theequation)\ltx@label{#1}&}
\makeatother
\allowdisplaybreaks

\begin{document}
%
\thispagestyle{empty}
\begin{flushright}
\end{flushright}
\vspace{0.8cm}

\begin{center}
{\Large\sc Towards the renormalisation of the Standard\\[0.4cm] Model
  effective field theory to dimension eight: \\[0.6cm] Bosonic interactions II
  }
\vspace{0.8cm}

\textbf{
S. Das Bakshi$^{\,a}$\footnote{\href{sdb@ugr.es}{sdb@ugr.es}}, M. Chala$^{\,a}$\footnote{\href{mikael.chala@ugr.es}{mikael.chala@ugr.es}}, A. D\'iaz-Carmona$^{\,a}$\footnote{\href{aldiaz@ugr.es}{aldiaz@ugr.es}} and G. Guedes$^{\,a,b}$\footnote{\href{gguedes@lip.pt}{gguedes@lip.pt}}}\\
\vspace{1.cm}
{\em {$^a$ CAFPE and Departamento de F\'isica Te\'orica y del Cosmos,
Universidad de Granada, Campus de Fuentenueva, E--18071 Granada, Spain}}\\[0.5cm]
{\em {$^b$ Laborat\'orio de Instrumenta\c cao e F\'isica Experimental de Part\'iculas, Departamento de
F\'isica da Universidade do Minho, Campus de Gualtar, 4710-057 Braga, Portugal}}\\[0.2cm]

\vspace{0.5cm}
\end{center}
\begin{abstract}
We calculate the renormalisation group running of the bosonic Standard Model (SM) effective operators at one loop and to order $v^4/\Lambda^4$, with $v\sim 246$ GeV being the electroweak scale and  $\Lambda$ the unknown new physics threshold. We focus on contributions driven by one dimension-eight term and SM couplings, thus extending (and completing) the effort initiated in Ref.~\cite{Chala:2021pll}, in which quantum corrections from pairs of dimension-six interactions were considered.
We highlight some interesting consequences, including the renormalisation of loop-induced interactions by tree-level generated terms and, more importantly, the validity of positivity bounds on different operators inducing anomalous gauge quartic couplings.
%
\end{abstract} 

\newpage

\tableofcontents

\section{Introduction}
The Standard Model (SM) effective field theory, commonly known as SMEFT~\cite{Brivio:2017vri}, is currently one of the most popular descriptions of the elementary particles and their interactions; mostly because it can correctly account for experimental observations not well explained within the SM alone without conflicting with the null results for resonant searches encountered at different experiments and in particular at the LHC. The appeal of the SMEFT also relies on the very mild assumptions on which it is built, namely the validity of the SM gauge symmetry $SU(3)_c\times SU(2)_L\times U(1)_Y$ and the presence of a mass gap between the electroweak (EW) scale and the new physics threshold. 

In the absence of lepton-number violation (LNV), the dimension-six terms of the SMEFT naively provide the dominant corrections to SM predictions in low-energy observables. In recent years though, the next tower of operators, namely those of dimension eight, are being more and more scrutinised both from the theory and 
experimental points of view~\cite{Alioli:2022fng}. One important reason for this is that, in a number of observables, the leading dimension-six contributions vanish~\cite{Azatov:2016sqh}. A more striking motivation is that dimension-eight operators are the first ones subject to the so-called positivity constraints, which are theoretical bounds on the signs of certain (combinations of) Wilson coefficients implied solely by the principles of unitarity and analyticity of the $S$-matrix~\cite{Adams:2006sv}. Thus, any experimental evidence of a violation of these constraints would indicate the invalidity of the EFT approach (for example, due to the existence of new light degrees of freedom) or even the breakdown of some of the fundamental principles of modern physics.

In a previous work~\cite{Chala:2021pll}, 
we started efforts to renormalise the SMEFT to order $v^4/\Lambda^4$ (with $v\sim 246$ GeV being the Higgs vacuum expectation value), thus including dimension-eight interactions. In that 
paper, the renormalisation group evolution of the bosonic sector of the SMEFT driven by pairs of dimension-six interactions at one-loop was computed. In this article, we focus on the renormalisation of the same set of operators but as triggered by dimension-eight terms. It is worth mentioning that, during the course of the work presented in this paper, a number of results related to the renormalisation of the dimension-eight SMEFT were presented in Ref.~\cite{AccettulliHuber:2021uoa}. These include the renormalisation group evolution (RGE) of both bosonic and fermionic operators, but restricted to linear order in the Higgs quartic parameter and to quadratic order in the gauge couplings. We do compute the higher-power corrections, which in particular induce (otherwise absent) mixing between several operators. Corrections to lower-dimensional operators, proportional to the Higgs squared mass, that we also include here, were disregarded in Ref.~\cite{AccettulliHuber:2021uoa} too.

This article is structured as follows. In section~\ref{sec:theory} we introduce our notation and provide details on the calculation procedure. In section~\ref{sec:rges} we discuss the structure of the anomalous dimension matrix, drawing special attention to
(i) elements that depart significantly from their naive power counting estimate; and (ii) interactions that, despite arising only at loop-level in renormalisable models of new physics, 
are renormalised by operators that can be generated at tree level. (We do not provide explicit expressions of all the RGEs in the text; they can be instead found in an auxiliary file on \href{github.com/SMEFT-Dimension8-RGEs}{https://github.com/SMEFT-Dimension8-RGEs}.)
In section~\ref{sec:positivity}, we discuss the behaviour of some positivity bounds under the quantum corrections derived in this paper.
We conclude in section~\ref{sec:conclusions}. We dedicate Appendix~\ref{app:comparison} to comparing our results with those obtained in Ref.~\cite{AccettulliHuber:2021uoa}.

\section{Theory and conventions}
\label{sec:theory}
The SMEFT Lagrangian is an expansion in inverse powers of the cutoff $\Lambda\gg v$. Assuming lepton-number conservation, it reads:
\begin{align}
 \mathcal{L}_\text{SMEFT} = \mathcal{L}_\text{SM} + \frac{1}{\Lambda^2}\sum_i c_i^{(6)} \mathcal{O}_i^{(6)} + \frac{1}{\Lambda^4}\sum_j c_j^{(8)}\mathcal{O}_j^{(8)} + \cdots 
\end{align}
where $\mathcal{L}_\text{SM}$ represents the SM dimension-four Lagrangian and, in our convention, $i$ and $j$ run over the operators in the bases of dimension-six and dimension-eight interactions given in Refs.~\cite{Grzadkowski:2010es} (the ``Warsaw'' basis) and \cite{Murphy:2020rsh}, respectively\footnote{Ref.~\cite{Li:2020gnx} also presented a basis of dimension-8 SMEFT operators concurrent to Ref.~\cite{Murphy:2020rsh}.}. We borrow the notation for the Wilson coefficients from this reference.
The ellipses encode operators of dimension ten and higher.

We write the renormalisable SM Lagrangian as follows: 
\begin{align}\nonumber
 \mathcal{L}_\text{SM} = & -\frac{1}{4}G_{\mu\nu}^{A}G^{A\,\mu\nu} -\frac{1}{4}W_{\mu\nu}^{a}W^{a\,\mu\nu} -\frac{1}{4}B_{\mu\nu}B^{\mu\nu}\\\nonumber
 &
+\overline{q_{L}^{\alpha}}\ii\slashed{D}q_{L}^{\alpha}
+\overline{l_{L}^{\alpha}}\ii\slashed{D}l_{L}^{\alpha}
+\overline{u_{R}^{\alpha}}\ii\slashed{D}u_{R}^{\alpha}
+\overline{d_{R}^{\alpha}}\ii\slashed{D}d_{R}^{\alpha}
+\overline{e_{R}^{\alpha}}\ii\slashed{D}e_{R}^{\alpha}
\\
& +\left(D_{\mu}\phi\right)^{\dagger}\left(D^{\mu}\phi\right)
+\mu^{2}|\phi|^{2}-\lambda|\phi|^{4}
-\left(
y_{\alpha\beta}^{u}\overline{q_{L}^{\alpha}}\widetilde{\phi}u_{R}^{\beta}
+y_{\alpha\beta}^{d}\overline{q_{L}^{\alpha}}\phi d_{R}^{\beta}
+y_{\alpha\beta}^{e}\overline{l_{L}^{\alpha}}\phi e_{R}^{\beta}
+\text{h.c.}\right)~.
\end{align}
We denote by $e$, $u$ and $d$ the right-handed leptons and
quarks; while $l$ and $q$ stand for the left-handed
counterparts. The letters $W, B$ and $G$ refer to the EW gauge bosons and the gluon, respectively. We represent the Higgs doublet by $\phi =  (\phi^+, \phi^0)^T$, and $\tilde{\phi} = \ii\sigma_2\phi^*$ with $\sigma_I$ ($I=1,2,3$) being the Pauli matrices. Our expression for the covariant derivative is:
\begin{equation}
 D_\mu = \partial_\mu - \ii g_1 Y B_\mu -ig_2\frac{\sigma^I}{2} W_\mu^I -\ii g_3\frac{\lambda^A}{2} G_\mu^A\,,
\end{equation}
where $g_1, g_2$ and $g_3$ represent, respectively, the $U(1)_Y$, $SU(2)_L$ and $SU(3)_c$ gauge couplings, $Y$ stands for the hypercharge and $\lambda^A$ are the Gell-Mann matrices.

To order $v^4/\Lambda^4$ 
and assuming lepton-number conservation, the dimension-eight Wilson coefficients are only renormalised by
dimension-eight couplings themselves, proportional to renormalisable terms, as well as by pairs of dimension-six
interactions. Schematically: 
\begin{equation}\label{eq:gprime}
\dot{c}_i^{(8)} \equiv 16\pi^2\tilde{\mu} \frac{d c_{i}^{(8)}}{d\tilde{\mu}} = \gamma_{ij} c_j^{(8)} + \gamma_{ijk}' c_j^{(6)}c_k^{(6)}\,. 
\end{equation}
The anomalous dimensions $\gamma_{ijk}'$ in the bosonic sector were computed in Ref.~\cite{Chala:2021pll}. In this work, we focus on the $\gamma_{ij}$ counterpart. 
The dimension-eight terms renormalise lower-dimensional interactions too, proportional to $\mu^2$. We also calculate these corrections in this work.

We carry renormalisation by computing the divergences of the operators in the basis of independent Green's functions of Ref.~\cite{Chala:2021cgt}, which extends the basis of independent physical operators of Ref.~\cite{Murphy:2020rsh}. The former operators can be projected onto the latter using the relations derived through the equations of motion in Ref.~\cite{Chala:2021cgt}. This procedure ensures that only off-shell 1-particle irreducible (1PI) Feynman diagrams have to be considered in the calculation.


For the computations of the diagrams, we use \texttt{FeynArts}~\cite{Hahn:2000kx} and \texttt{FormCalc}~\cite{Hahn:1998yk} using the \texttt{Feynrules}~\cite{Alloul:2013bka} model provided in Ref.~\cite{Chala:2021cgt}, too. We work in dimensional regularisation with space-time dimension $\text{d}=4-2\epsilon$, using the background field method and the Feynman gauge. As a matter of example, we provide below a detailed computation of the mixing of $\mathcal{O}_{B\phi^4 D^2}^{(1)}$ into the three operators in the class $\phi^4 D^4$, showing explicitly that it vanishes.

%
Following the results of Ref.~\cite{Chala:2021cgt}, the interactions $\mathcal{O}_{\phi^4 D^4}^{(1,2,3)}$ receive contributions from different redundant operators: 
\begin{align}\label{eq:wcshiftexample}
c_{\phi^4}^{(1)}&\rightarrow c_{\phi^4}^{(1)} + g_1^2 c_{B^2D^4} - g_1 c_{B\phi^2D^4}^{(3)} - g_2^2 c_{W^2D^4}+g_2 c_{W\phi^2D^4}^{(3)}\,,\nonumber\\
c_{\phi^4}^{(2)}&\rightarrow c_{\phi^4}^{(2)} - g_1^2 c_{B^2D^4} + g_1 c_{B\phi^2D^4}^{(3)} - g_2^2 c_{W^2D^4}+g_2 c_{W\phi^2D^4}^{(3)}\,,\nonumber\\
c_{\phi^4}^{(3)}&\rightarrow c_{\phi^4}^{(3)} + 2g_2^2 c_{W^2D^4} - 2g_2 c_{W\phi^2D^4}^{(3)}\,.
\end{align}
Thus, one needs to compute the divergences of these operators generated by loops involving one insertion of $\mathcal{O}_{B\phi^4 D^2}^{(1)}$. We note, however, that this insertion does not generate one-loop diagrams with only one $W$ and two Higgses as external particles, or with two external $B$ (or $W$) bosons and no Higgses. This will be seen in more detail in the next section. Moreover, those diagrams renormalising $\mathcal{O}_{B\phi^2 D^4}^{(3)}$ necessarily involve a Higgs bubble, the divergence of which is proportional to $\mu^2/\Lambda^4$ and hence of dimension six. Therefore, the only relevant divergences are those of the operators $\phi^4 D^4$ themselves.

\begin{figure}
\begin{center}
\includegraphics[scale=1.0]{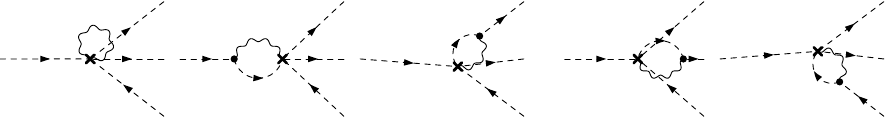}
\end{center}
\caption{1-loop diagrams for the process $\phi\rightarrow\phi^\dagger\phi\phi$. Dashed lines represent the Higgs doublet, wavy lines represent $B$ bosons and crosses represent insertions of $\mathcal{O}_{B\phi^4 D^2}^{(1)}$.}\label{fig:example1}
\end{figure}

In order to compute these, we calculate the 1PI one-loop amplitude $\phi\rightarrow\phi^\dagger \phi \phi$; the corresponding Feynman diagrams are obtained with \texttt{FeynArts} and are shown in Fig.~\ref{fig:example1}. The computation of the amplitude is performed using \texttt{FormCalc} and the divergent part of the result reads:
\begin{align}
-i\mathcal{A}_{UV}=\frac{1}{64\pi^2\epsilon}c_{B\phi^4D^2}^{(1)}&\bigg[2(\kappa_{3333}-\kappa_{1233}-2\kappa_{1323}-2\kappa_{1333}-2\kappa_{2333}+\kappa_{1313}+\kappa_{2323})\nonumber\\\nonumber
&+\kappa_{1133} +\kappa_{1123}+\kappa_{1213}+\kappa_{2233}+\kappa_{1112}\\
&-\kappa_{1113}+\kappa_{1222}-\kappa_{1223}-\kappa_{1322}+\kappa_{2223}\bigg]\,,
\end{align}
with $\kappa_{ijkl}=(p_i\cdot p_j) (p_k \cdot p_l)$. This divergence must be absorbed by local counterterms. In order to determine their values, we compute the same $\phi\to\phi^\dagger\phi\phi$ amplitude at tree-level in the EFT using again \texttt{FormCalc}. The result is given by:
\begin{align}
-i\mathcal{A}_{IR}&=2c_{\phi^4}^{(1)}\left(-\kappa_{1213}+\kappa_{1322}+\kappa_{1323}\right)+2c_{\phi^4}^{(2)}\left(-\kappa_{1123}+\kappa_{1223}+\kappa_{1323}\right)\nonumber\\
&+2c_{\phi^4}^{(3)}\left(-\kappa_{1213}+\kappa_{1223}+\kappa_{1233}\right)+2c_{\phi^4}^{(4)}\left(\kappa_{1212}+\kappa_{1213}-\kappa_{1223}-\kappa_{1233}\right)\nonumber\\
&+c_{\phi^4}^{(4)}\left(-\kappa_{1112}+\kappa_{1113}-\kappa_{1123}-\kappa_{1133}-\kappa_{1222}+\kappa_{1322}-\kappa_{2223}-\kappa_{2233}\right)\nonumber\\
&+c_{\phi^4}^{(6)}\left(\kappa_{1112}-\kappa_{1113}-\kappa_{1122}-\kappa_{1123}+\kappa_{1233}-\kappa_{1333}-\kappa_{2223}-\kappa_{2333}\right)\nonumber\\
&+2c_{\phi^4}^{(6)}\left(\kappa_{1313}+\kappa_{1213}-\kappa_{1322}-\kappa_{1323}\right)+4c_{\phi^4}^{(8)}\left(-\kappa_{1112}-\kappa_{1113}+\kappa_{1123}\right)\nonumber\\
&+2c_{\phi^4}^{(8)}\left(\kappa_{1111}+\kappa_{1122}+\kappa_{1133}+\kappa_{2233}\right)+c_{\phi^4}^{(10)}\left(\kappa_{3333}+\kappa_{1122}+\kappa_{1133}+\kappa_{2233}\right)\nonumber\\
&+2c_{\phi^4}^{(10)}\left(-\kappa_{1233}-\kappa_{1333}+\kappa_{2333}\right)+c_{\phi^4}^{(11)}\left(\kappa_{2222}+\kappa_{1122}+\kappa_{1133}+\kappa_{2333}\right)\nonumber\\
&+2c_{\phi^4}^{(11)}\left(-\kappa_{1222}-\kappa_{1322}+\kappa_{2223}\right)+2c_{\phi^4}^{(12)}\left(\kappa_{2323}+\kappa_{1123}-\kappa_{1223}-\kappa_{1323}\right)\nonumber\\&+c_{\phi^4}^{(12)}\left(-\kappa_{1122}-\kappa_{1133}+\kappa_{1222}+\kappa_{1233}+\kappa_{1322}+\kappa_{1333}+\kappa_{2223}+\kappa_{2333}\right)
\end{align}
(There are also CP-violating counterterms that we ignore in this expression because they are not relevant to absorb the divergences from the insertion of the CP-conserving $\mathcal{O}_{B\phi^4 D^2}^{(1)}$).
After equating $\mathcal{A}_{UV}$ and $\mathcal{A}_{IR}$, while applying momentum conservation, we obtain:
%
\begin{align}
c_{\phi^4}^{(6)}=\frac{g_2}{64\pi^2\epsilon}c_{B\phi^4D^2}^{(1)}\,,\quad c_{\phi^4}^{(10)}=\frac{g_2}{32\pi^2\epsilon}c_{B\phi^4D^2}^{(1)}\,,\quad c_{\phi^4}^{(12)}&=\frac{g_2}{32\pi^2\epsilon}c_{B\phi^4D^2}^{(1)}\,,
\end{align}
with all other counterterms being zero. These particular $\phi^4 D^4$ operators are redundant and do not contribute to $\mathcal{O}_{\phi^4 D^4}^{(1,2,3)}$ on-shell as indicated in  Eq.~\eqref{eq:wcshiftexample} (in fact they only contribute to operators in the $\phi^6 D^2$ and $\phi^8$ classes; see Ref.~\cite{Chala:2021cgt}). Therefore, we conclude that the interactions $\mathcal{O}_{\phi^4 D^2}^{(1,2,3)}$ are not renormalised by $\mathcal{O}_{B\phi^4 D^2}^{(1)}$.

We have cross checked most of our results (with perfect agreement in all cases) with the help of \texttt{matchmakereft}~\cite{Carmona:2021xtq}. In fact, for the calculation of eight-Higgs processes we have relied entirely on this latter tool. We neglect loops with insertions of operators that can arise only at loop-level in weakly-coupled UV completions of the SMEFT~\cite{Craig:2019wmo}, as these effects are formally two-loop corrections.

\section{Structure of the anomalous dimension matrix}
\label{sec:rges}
\begin{table}[t]
 \begin{center}
 \resizebox{0.9\textwidth}{!}{\begin{tabular}{c|ccccccccc}
  \toprule
  & $\phi^4 D^4$ & $B\phi^4 D^2$ & $W\phi^4 D^2$ & $B^2\phi^4$ & $W^2\phi^4$ & $WB\phi^4$ & $G^2\phi^4$ & $\phi^6 D^2$ & $\phi^8$\\\midrule
  \textcolor{gray}{$B^2 \phi^2 D^2$} & \graycell$g_1^2$ & 0 & 0 & 0 & 0 & 0 & 0 & 0 & 0 \\[0.1cm]
  \textcolor{gray}{$W^2 \phi^2 D^2$} & \graycell$g_2^2$ & 0 & 0 & 0 & 0 & 0 & 0 & 0 & 0 \\[0.1cm]
  \textcolor{gray}{$WB \phi^2 D^2$} & \graycell$g_1 g_2$ & 0 & 0 & 0 & 0 & 0 & 0 & 0 & 0 \\[0.1cm]
  \textcolor{gray}{$G^2 \phi^2 D^2$} & 0 & 0 & 0 & 0 & 0 & 0 & 0 & 0 & 0 \\[0.1cm]
  \textcolor{gray}{$W^3 \phi^2$} & 0 & 0 & 0 & 0 & 0 & 0 & 0 & 0 & 0 \\[0.1cm]
  \textcolor{gray}{$W^2 B \phi^2$} & 0 & 0 & 0 & 0 & 0 & 0 & 0 & 0 & 0 \\[0.1cm]
  \textcolor{gray}{$G^3 \phi^2$} & 0 & 0 & 0 & 0 & 0 & 0 & 0 & 0 & 0 \\[0.1cm]
  $\phi^4 D^4$ & $\mrr{g_2^2}$ & 0 & 0 & 0 & 0 & 0 & 0 & 0 & 0\\[0.1cm]
  $B\phi^4 D^2$ & $g_1g_2^2$ & $\mrr{\lambda}$ & 0 & 0 & 0 & 0 & 0 & 0 & 0 \\[0.1cm]
  $W\phi^4 D^2$ & $g_2^3$ & 0 & $g_2^2$ & 0 & 0 & 0 & 0 & 0 & 0 \\[0.1cm]
  $B^2\phi^4$ & $g_1^2 g_2^2$ & $g_1 \lambda$ & $g_1^2 g_2$ & $\mrr{\lambda}$ & 0 & $g_1 g_2$ & 0 & 0 & 0 \\[0.1cm]
  $W^2\phi^4$ & $g_2^4$ & $g_1 g_2^2$ & $g_2^3$ & 0 & $\mrr{\lambda}$ & $g_1 g_2$ & 0 & 0 & 0 \\[0.1cm]
  $WB\phi^4$ & $g_1 g_2^3$ & $g_2 \lambda$ & $g_1 \lambda$ & $g_1 g_2$ & $g_1 g_2$ & $\mrr{\lambda}$ & 0 & 0 & 0 \\[0.1cm]
  $G^2\phi^4$ & 0 & 0 & 0 & 0 & 0 & 0 & $\mrr{g_3^2}$ & 0 & 0 \\[0.1cm]
  $\phi^6 D^2$ & $\mrr{g_2^4}$ & $ \mrr{g_1 \lambda}$ & $\mrr{g_2 \lambda}$ & 0 & 0 & 0 & 0 & $\mrr{\lambda}$ & 0 \\[0.1cm]
  $\phi^8$ & $\mrr{\lambda^3}$ & $\mrr{g_1\lambda^2}$ & $\mrr{g_2\lambda^2}$ & $\mrr{g_1^2 \lambda}$ & $\mrr{g_2^2 \lambda}$ & $\mrr{g_1 g_2\lambda}$ & 0 & $\mrr{\lambda^2}$ & $\mrr{\lambda}$\\[0.1cm]
  \bottomrule
 \end{tabular}
 }
 \end{center}
 \caption{\it Structure of the bosonic-bosonic dimension-eight anomalous dimension matrix. The entries indicate the order in SM couplings of the leading contribution. Those in blue represent terms that deviate significantly from naive dimensional analysis. The operators in gray can only arise at loop-level in weakly-coupled UV completions of the SMEFT. The shaded cells indicate those of the latter operators that are renormalised by interactions that can be generated at tree level.}\label{tab:dim8adm}
\end{table}
\begin{table}[t]
 \begin{center}
 \resizebox{\textwidth}{!}{\begin{tabular}{c|cccccccccc}
  \toprule
  & $\psi^2 B \phi^3$ & $\psi^2 W \phi^3$ & $\psi^2 G \phi^3$ & $\psi^2 \phi^2 D^3$ & $\psi^2\phi^5$ & $\psi^2\phi^4 D$ & $\psi^2 B\phi^2 D$ & $\psi^2 W\phi^2 D$ & $\psi^2 G\phi^2 D$ & $\psi^2 \phi^3 D^2$\\\midrule
  \textcolor{gray}{$B^2 \phi^2 D^2$} & 0 & 0 & 0 & \graycell $ g_1^2 $ & 0 & 0 & 0 & 0 & 0 & 0\\[0.1cm]
  \textcolor{gray}{$W^2 \phi^2 D^2$} & 0 & 0 & 0 & \graycell $ g_2^2 $ & 0 & 0 & 0 & 0 & 0 & 0\\[0.1cm]
  \textcolor{gray}{$WB \phi^2 D^2$} & 0 & 0 & 0 & \graycell $ g_1 g_2 $ & 0 & 0 & 0 & 0 & 0 & 0\\[0.1cm]
  \textcolor{gray}{$G^2 \phi^2 D^2$} & 0 & 0 & 0 & \graycell $ g_3^2 $ & 0 & 0 & 0 & 0 & 0 & 0\\[0.1cm]
  \textcolor{gray}{$W^3 \phi^2$} & 0 & 0 & 0 & 0 & 0 & 0 & 0 & 0 & 0 & 0\\[0.1cm]
  \textcolor{gray}{$W^2 B \phi^2$} & 0 & 0 & 0 & 0 & 0 & 0 & 0 & 0 & 0 & 0\\[0.1cm]
  \textcolor{gray}{$G^3 \phi^2$} & 0 & 0 & 0 & 0 & 0 & 0 & 0 & 0 & 0 & 0\\[0.1cm]
  $\phi^4 D^4$ & 0 & 0 & 0 & \mrr{$ |y^t|^2 $} & 0 & 0 & 0 & 0 & 0 & 0\\[0.1cm]
  $B\phi^4 D^2$ & 0 & 0 & 0 & \mrr{$g_1 |y^t|^2 $} & 0 & 0 & \mrr{$ |y^t|^2 $} & 0 & 0 & $ g_1 y^t $\\[0.1cm]
  $W\phi^4 D^2$ & 0 & 0 & 0 & \mrr{$g_2 |y^t|^2 $} & 0 & 0 & 0 & \mrr{$ |y^t|^2 $} & 0 & $ g_2 y^t $\\[0.1cm]
  $B^2\phi^4$ & $ g_1 y^t $ & 0 & 0 & $g_1^2 |y^t|^2 $ & 0 & 0 & $ g_1 |y^t|^2 $ & 0 & 0 & $ g_1^2 y^t $\\[0.1cm]
  $W^2\phi^4$ & 0 & $ g_2 y^t $ & 0 & $g_2^2 |y^t|^2 $ & 0 & 0 & 0 & $ g_2 |y^t|^2 $ & 0 & $ g_2^2 y^t $\\[0.1cm]
  $WB\phi^4$ & $ g_2 y^t $& $ g_1 y^t $ & 0 & $g_1 g_2 |y^t|^2 $& 0 & 0 & $ g_2 |y^t|^2 $ & $ g_1 |y^t|^2 $ & 0 & $ g_1 g_2 y^t $\\[0.1cm]
  $G^2\phi^4$ & 0 & 0 & $ g_3 y^t $ & 0 & 0 & 0 & 0 & 0 & 0 & 0\\[0.1cm]
  $\phi^6 D^2$ & 0 & 0 & 0 & \mrr{$g_2^2 |y^t|^2 $} & 0 & \mrr{$ |y^t|^2 $} & $ g_1 |y^t|^2 $ & \mrr{$ g_2 |y^t|^2 $} & 0 & \mrr{$y^t |y^t|^2$ } \\[0.1cm]
  $\phi^8$ & 0 & 0 & 0 & \mrr{$ \lambda |y^t|^4 $} & \mrr{$ y^t |y^t|^2 $} & \mrr{$ \lambda |y^t|^2 $} & $ g_1 \lambda |y^t|^2 $ & \mrr{$g_2 \lambda |y^t|^2 $} & 0 & \mrr{$\lambda y^t |y^t|^2$}\\[0.1cm]
  \bottomrule
 \end{tabular}
 }
 \end{center}
 \caption{\it Same as Tab.~\ref{tab:dim8adm} but for the bosonic-fermionic anomalous dimension matrix.}\label{tab:dim8adm2}
\end{table}
\begin{table}[t]
 \begin{center}
 \resizebox{0.9\textwidth}{!}{\begin{tabular}{c|cccccccccccc}
  \toprule
  & $\phi^4 D^4$ & $B\phi^4 D^2$ & $W\phi^4 D^2$ & $B^2\phi^4$ & $W^2\phi^4$ & $WB\phi^4$ & $G^2\phi^4$ & $\phi^6 D^2$ & $\phi^8$\\\midrule
  $\phi^2$ & $ \mu^6 $ & 0 & 0 & 0 & 0 & 0 & 0 & 0 & 0 \\[0.1cm]
  %
  %
  $\phi^4$ & $ \mrr{\lambda \mu^4} $ & $ g_1 \mu^4 $ & $ \mrr{g_2 \mu^4} $ & 0 & 0 & 0 & 0 & $ \mu^4 $ & 0 \\\midrule
  %
  %
  %
  %
  %
  \textcolor{gray}{$B^2 \phi^2$} & \graycell$g_1^2 \mu^2$ & \graycell$g_1 \mu^2$ & 0 & \graycell$\mrr{\mu^2}$ & 0 & 0 & 0 & 0 & 0\\[0.1cm]
  \textcolor{gray}{$W^2 \phi^2$} & \graycell$g_2^2\mu^2$ & 0 & \graycell$g_2\mu^2$ & 0 & \graycell$\mrr{\mu^2}$ & 0 & 0 & 0 & 0\\[0.1cm]
  \textcolor{gray}{$WB \phi^2$} & \graycell$g_1g_2\mu^2$ & \graycell$g_2\mu^2$ & \graycell$g_1\mu^2$ & 0 & 0 & \graycell $\mu^2$ & 0 & 0 & 0\\[0.1cm]
  \textcolor{gray}{$G^2 \phi^2$} & 0 & 0 & 0 & 0 & 0 & 0 & \graycell$\mrr{\mu^2}$ & 0 & 0\\[0.1cm]
  $\phi^4 D^2$ & $\mrr{\lambda\mu^2}$ & $g_1 \mu^2$ & $ \mrr{g_2 \mu^2} $ & 0 & 0 & 0 & 0 & $\mrr{\mu^2}$ & 0\\[0.1cm]
  $\phi^6$ & $\mrr{\lambda^2\mu^2}$ & $\mrr{\lambda g_1\mu^2}$ & $\mrr{\lambda g_2\mu^2}$ & $g_1^2\mu^2$ &  $\mrr{g_2^2\mu^2}$ & $g_1 g_2\mu^2$ & 0 & $\mrr{\lambda\mu^2}$ & $\mrr{\mu^2}$\\[0.1cm]
  \bottomrule
 \end{tabular} 
 }
 \end{center}
 \caption{\it Same as Tab.~\ref{tab:dim8adm} but for the renormalisation of the bosonic interactions of dimensions two, four and six.}\label{tab:dim6adm}
\end{table}
\begin{table}[t]
 \begin{center}
 \resizebox{\textwidth}{!}{\begin{tabular}{c|ccccccccccccc}
  \toprule
  & $\psi^2 B \phi^3$ & $\psi^2 W \phi^3$ & $\psi^2 G \phi^3$ & $\psi^2 \phi^2 D^3$ & $\psi^2\phi^5$ & $\psi^2\phi^4 D$ & $\psi^2 B\phi^2 D$ & $\psi^2 W\phi^2 D$ & $\psi^2 G\phi^2 D$ & $\psi^2 \phi^3 D^2$\\\midrule
  $\phi^2$ & 0 & 0 & 0 & 0 & 0 & 0 & 0 & 0 & 0 & 0\\[0.1cm]
  $\phi^4$ & 0  & 0 & 0 & \mrr{$ \mu^4 |y^t|^2 $} & 0 & 0 & 0 & 0 & 0 & $ \mu^4 y^t $\\\midrule
  \textcolor{gray}{$B^2 \phi^2$} & 0 & 0 & 0 & 0 & 0 & 0 & 0 & 0 & 0 & 0\\[0.1cm]
  \textcolor{gray}{$W^2 \phi^2$} & 0 & 0 & 0 & 0 & 0 & 0 & 0 & 0 & 0 & 0\\[0.1cm]
  \textcolor{gray}{$WB \phi^2$} & 0 & 0 & 0 & 0 & 0 & 0 & 0 & 0 & 0 & 0\\[0.1cm]
  \textcolor{gray}{$G^2 \phi^2$} & 0 & 0 & 0 & 0 & 0 & 0 & 0 & 0 & 0 & 0\\[0.1cm]
  $\phi^4 D^2$ & 0 & 0 & 0 & \mrr{$ \mu^2 |y^t|^2 $} & 0 & 0 & 0 & 0 & 0 & $ \mu^2 y^t $\\[0.1cm]
  $\phi^6$ & 0 & 0 & 0 & \mrr{$ \lambda \mu^2 |y^t|^2 $} & $ \mu^2 y^t $ & \mrr{$\mu^2 |y^t|^2$} & $\mu^2 |y^t|^2$ & $\mu^2 |y^t|^2$ & 0 & $ \mu^2 y^t |y^t|^2  $\\[0.1cm]
  \bottomrule
 \end{tabular} 
 }
 \end{center}
 \caption{\it Same as Tab.~\ref{tab:dim8adm2} but for the renormalisation of the bosonic interactions of dimensions two, four and six.}\label{tab:dim6adm2}
\end{table}
The $\mathcal{O}(v^4/\Lambda^4)$ piece of the RGEs of both relevant and marginal bosonic interactions of the SMEFT, driven by operators that can arise at tree level upon matching, are provided in an auxiliary file on \href{github.com/SMEFT-Dimension8-RGEs}{https://github.com/SMEFT-Dimension8-RGEs}.
Here we limit ourselves to discussing some generic aspects of this result. 

\begin{figure}[!tbp]
  \centering
  \subfloat[]{\includegraphics[width=0.3\textwidth]{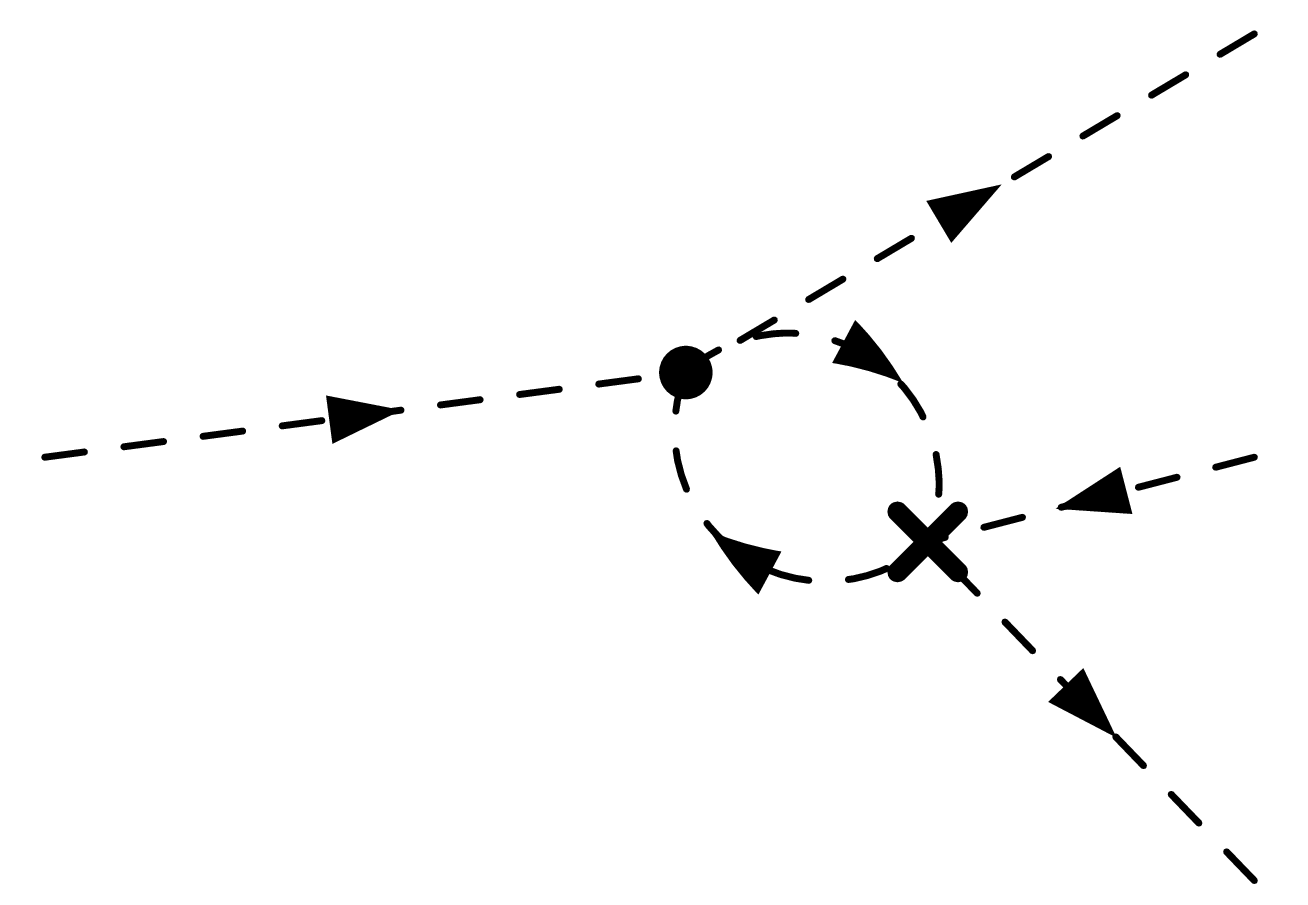}\label{fig:bosonic}}
  \hspace{1.5cm}
  \subfloat[]{\includegraphics[width=0.3\textwidth]{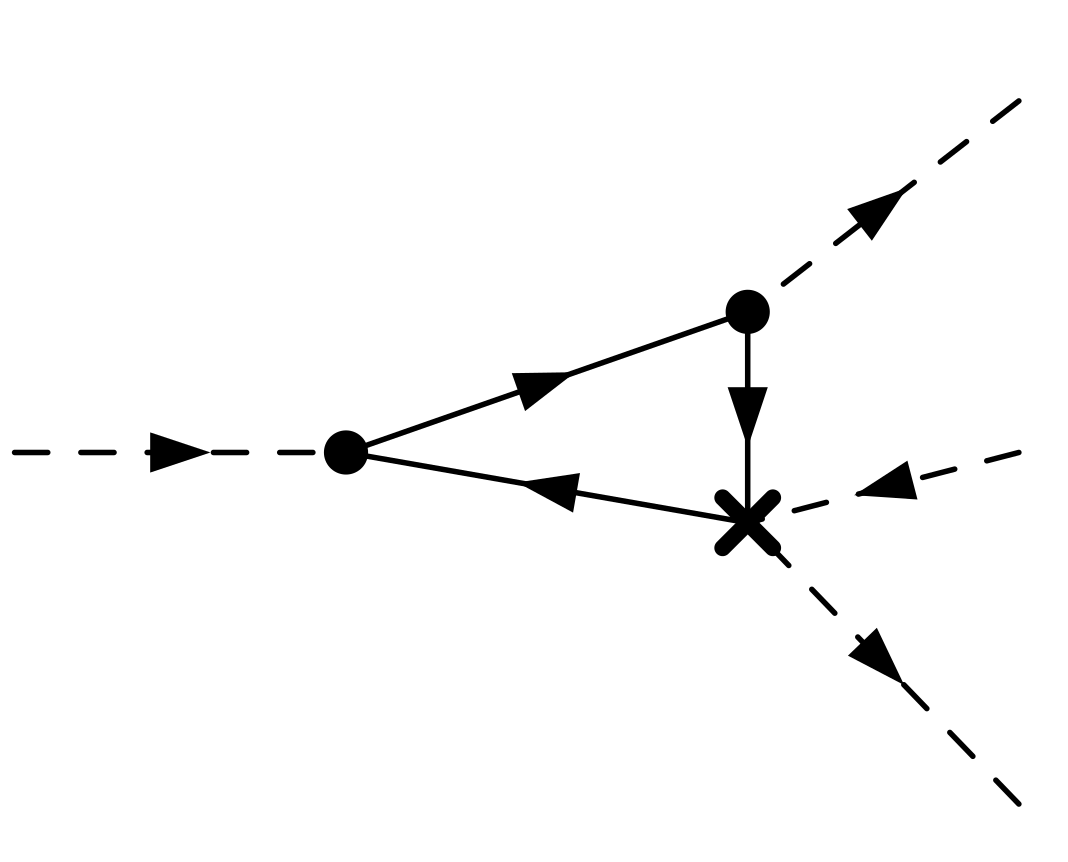}\label{fig:fermionic}}
  \caption{{\it Examples of diagrams contributing to the renormalisation of four-Higgs operators through insertions of (a) four-Higgs terms (b) two-fermion terms. The cross corresponds to a dimension-eight vertex.}}
\end{figure}

The structure of the anomalous dimension matrix $\gamma$ of dimension-eight bosonic operators is shown in Tabs.~\ref{tab:dim8adm} and ~\ref{tab:dim8adm2} for the contributions resulting from the insertion of bosonic and fermionic dimension-eight terms respectively. The entries of the tables correspond to the order in the SM couplings of the leading contribution of the renormalisation of the operators in the rows by insertions of the dimension-eight operators present in the columns. For the latter operators, we considered only those which can be generated at tree-level by weakly-coupled UV theories~\cite{Craig:2019wmo} which can renormalise bosonic operators. Note that since we are computing the RGEs at one-loop, at most two of the fields composing the inserted operator can be taken as internal particles, while the remaining ones will be external. Thus, since tree-level generated bosonic interactions involve at least four Higgses~\cite{Craig:2019wmo}, these will not renormalise operators that only involve gauge bosons which are therefore not shown in the tables. This is illustrated in Fig. \ref{fig:bosonic}, where we see an example of an insertion of a four-Higgs term that contributes to processes with external Higgses.

Furthermore, for the fermionic contributions, since terms with more than two fermions do not renormalise bosonic operators (because at most two particles can be taken internal in a given vertex), we considered only insertions of two-fermion operators in Tab.~\ref{tab:dim8adm2}. An example of this is shown in Fig. \ref{fig:fermionic}, where the two fermions composing the dimension-eight insertion are inside the loop.

%

There are mixing terms of a size that deviate significantly from the naive dimensional analysis estimate, $\gamma_{ij}/g\sim 1$; with $g$ representing (products of) SM couplings. In particular, we highlight in blue those that fulfil $\gamma_{ij}/g\gtrsim 10$. The corresponding RGEs read:
\begin{align}
 \dot{c}_{\phi^4}^{(3)} &= -g_2^2(12 c_{\phi^4}^{(1)} + \frac{29}{3}c_{\phi^4}^{(2)} + 14 c_{\phi^4}^{(3)})  - 56 (c^{(4)}_{q^2 \phi^2 D^3})_{_{\alpha_1,\alpha_2}} \, y^u_{\alpha_2,\alpha_3} (y^u)^*_{\alpha_3,\alpha_1}  + \cdots\\[0.1cm]
 \dot{c}_{B\phi^4D^2}^{(1)} &= 12 \lambda  c_{B\phi^4D^2}^{(1)} + 60 g_1 (c^{(4)}_{q^2 \phi^2 D^3})_{_{\alpha_1,\alpha_2}} \, y^u_{_{\alpha_2,\alpha_3}} (y^u)^*_{_{\alpha_3,\alpha_1}}\nonumber\\
 & -36 (c^{(1)}_{q^2 B \phi^2 D})_{_{\alpha_1,\alpha_2}} \, y^u_{_{\alpha_2,\alpha_3}} (y^u)^*_{_{\alpha_1,\alpha_3}} + \cdots\\[0.1cm]
 \dot{c}_{W\phi^4D^2}^{(1)} &= 44 g_2  (c^{(4)}_{q^2 \phi^2 D^3})_{_{\alpha_1,\alpha_2}} \, y^u_{_{\alpha_2,\alpha_3}} (y^u)^*_{_{\alpha_3,\alpha_1}} + 48 (c^{(11)}_{q^2 W H^2 D})_{_{\alpha_1,\alpha_2}} \, y^u_{_{\alpha_2,\alpha_3}} (y^u)^*_{_{\alpha_1,\alpha_3}} + \cdots\\[0.1cm]
 \dot{c}_{B^2\phi^4}^{(i)} &= 48 \lambda  c_{B^2\phi^4}^{(i)} + \cdots\\[0.1cm]
 \dot{c}_{W^2\phi^4}^{(i)} &= 48 \lambda c_{W^2\phi^4}^{(i)}+\cdots\\[0.1cm]
 \dot{c}_{W^2\phi^4}^{(j)} &= 24 \lambda c_{W^2\phi^4}^{(j)}+\cdots\\[0.1cm]
 \dot{c}_{WB\phi^4}^{(i)} &= 40 \lambda c_{WB\phi^4}^{(i)}+\cdots\\[0.1cm]
 \dot{c}_{G^2\phi^4}^{(i)} &= -14 g_3^2 c_{G^2\phi^4}^{(i)}+\cdots\\[0.1cm]
 \dot{c}_{\phi^6}^{(1)} &= -\frac{157}{16}g_2^4 c_{\phi^4 D^4}^{(3)} + 20 g_1 \lambda c_{B\phi^4D^2}^{(1)} + 40 g_2 \lambda c_{W\phi^4D^2}^{(1)} + 68\lambda c_{\phi^6}^{(1)} \nonumber \\[0.1cm] 
 &- 34 g_2^2 (c^{(4)}_{q^2 \phi^2 D^3})_{_{\alpha_1,\alpha_2}} \, y^u_{_{\alpha_2,\alpha_3}} (y^u)^*_{_{\alpha_3,\alpha_1}} - 48 (c^{(2)}_{q^2 \phi^4 D})_{_{\alpha_2,\alpha_1}}  \, y^u_{_{\alpha_1,\alpha_3}} (y^u)^*_{_{\alpha_2,\alpha_3}}  \nonumber \\[0.1cm]
 &-30 g_2 (c^{(11)}_{q^2 W \phi^2 D})_{_{\alpha_1,\alpha_2}}   \, y^u_{_{\alpha_2,\alpha_3}} (y^u)^*_{_{\alpha_1,\alpha_3}} -12 (c^{(1)}_{qu \phi^3 D^2})_{_{\alpha_1,\alpha_2}}   \, y^u_{_{\alpha_4,\alpha_3}} (y^u)^*_{_{\alpha_1,\alpha_3}} (y^u)^*_{_{\alpha_4,\alpha_2}}  +\cdots   \\[0.1cm]
 \dot{c}_{\phi^8} &= \frac{184}{3}\lambda^3 c_{\phi^4}^{(2)} -12 g_1\lambda^2 c_{B\phi^4 D^2}^{(1)}-16g_2\lambda^2 c_{W\phi^4 D^2}^{(1)}+12 g_1^2\lambda c_{B^2\phi^4}^{(1)} \nonumber \\
 &+ 36 g_2^2 \lambda c_{W^2\phi^4}^{(1)} + 12 g_1 g_2\lambda c_{WB\phi^4}^{(1)}+48 \lambda^2 c_{\phi^6 D^2}^{(2)}+192\lambda c_{\phi^8} \nonumber \\
 & +24\lambda (c^{(1)}_{u^2 \phi^2 D^3} +c^{(2)}_{u^2 \phi^2 D^3} )_{_{\alpha_1,\alpha_2}} \, y^u_{_{\alpha_3,\alpha_4}} y^u_{_{\alpha_5,\alpha_1}} (y^u)^*_{_{\alpha_2,\alpha_3}}  (y^u)^*_{_{\alpha_4,\alpha_5}} \nonumber \\
 & +24\lambda (c^{(1)}_{q^2 \phi^2 D^3} +c^{(2)}_{q^2 \phi^2 D^3} -c^{(3)}_{q^2 \phi^2 D^3} -c^{(4)}_{q^2 \phi^2 D^3} )_{_{\alpha_1,\alpha_2}} \, y^u_{_{\alpha_2,\alpha_3}} y^u_{_{\alpha_4,\alpha_5}} (y^u)^*_{_{\alpha_3,\alpha_4}}  (y^u)^*_{_{\alpha_5,\alpha_1}}\nonumber \\
 & -12  (c_{qu \phi^5})_{_{\alpha_1,\alpha_2}}  \, y^u_{_{\alpha_3,\alpha_4}} (y^u)^*_{_{\alpha_1,\alpha_4}} (y^u)^*_{_{\alpha_3,\alpha_2}} -12  (c_{qu \phi^5})^{*}_{_{\alpha_1,\alpha_2}}  \, y^u_{_{\alpha_1,\alpha_3}} (y^u)_{_{\alpha_4,\alpha_2}} (y^u)^*_{_{\alpha_4,\alpha_3}}\nonumber\\
 & +24\lambda (c^{(2)}_{q^2 \phi^4 D})_{_{\alpha_2,\alpha_1}} \, y^u_{_{\alpha_1,\alpha_3}} (y^u)^*_{_{\alpha_2,\alpha_3}}    +12 g_2 \lambda (c^{(11)}_{q^2 W \phi^2 D})_{_{\alpha_1,\alpha_2}}   \, y^u_{_{\alpha_2,\alpha_3}} (y^u)^*_{_{\alpha_1,\alpha_3}}\nonumber\\
 & -12 \lambda (c^{(5)}_{qu \phi^3 D^2})_{_{\alpha_1,\alpha_2}}   \, y^u_{_{\alpha_4,\alpha_3}} (y^u)^*_{_{\alpha_1,\alpha_3}} (y^u)^*_{_{\alpha_4,\alpha_2}} + \cdots
\end{align}
with $i=1,2$ and $j=3,4$; $\alpha_i$'s are the flavor indices.

The %
shaded cells represent operators (those of the form $X^2\phi^2 D^2$) that, despite arising only at the loop-level in weakly-coupled UV completions of the SMEFT, are renormalised by interactions that can be generated at tree level (those of the form $\phi^4 D^4$ and $\psi^2 \phi^2 D^3$). (This is in contrast with what occurs within the dimension-six bosonic sector of the SMEFT~\cite{Cheung:2015aba}.) The RGEs of these operators read: 
\begin{align}\label{eq:treeloppmixing}
 \dot{c}_{B^2\phi^2 D^2}^{(1)} &= \frac{1}{6} g_1^2 (2 c_{\phi^4}^{(1)} + 3 c_{\phi^4}^{(2)} + c_{\phi^4}^{(3)}) -\frac{8}{9}g_1^2 \left[ 2 c^{(1)}_{d^2 \phi^2 D^3} + 2 c^{(2)}_{d^2 \phi^2 D^3} + 8( c^{(1)}_{u^2 \phi^2 D^3} + c^{(2)}_{u^2 \phi^2 D^3}) \right. \,\nonumber\\
 & \left. +6 c^{(1)}_{e^2 \phi^2 D^3} + 3 c^{(1)}_{l^2 \phi^2 D^3} + 3 c^{(2)}_{l^2 \phi^2 D^3} + c^{(1)}_{q^2 \phi^2 D^3} + c^{(2)}_{q^2 \phi^2 D^3}  \right]_{\alpha_1,\alpha_1}\,,\\
 \dot{c}_{B^2\phi^2 D^2}^{(2)} &= -\frac{1}{24} g_1^2 (2 c_{\phi^4}^{(1)} + 3 c_{\phi^4}^{(2)} + c_{\phi^4}^{(3)})\,+\frac{2}{9}g_1^2 \left[ 2 c^{(1)}_{d^2 \phi^2 D^3} + 2 c^{(2)}_{d^2 \phi^2 D^3} + 8( c^{(1)}_{u^2 \phi^2 D^3} + c^{(2)}_{u^2 \phi^2 D^3}) \right. \,\nonumber\\
 & \left. +6 c^{(1)}_{e^2 \phi^2 D^3} + 3 c^{(1)}_{l^2 \phi^2 D^3} + 3 c^{(2)}_{l^2 \phi^2 D^3} + c^{(1)}_{q^2 \phi^2 D^3} + c^{(2)}_{q^2 \phi^2 D^3}  \right]_{\alpha_1,\alpha_1}  \,,\\[0.3cm]
 \dot{c}_{W^2\phi^2 D^2}^{(1)} &= \frac{1}{6} g_2^2 (2 c_{\phi^4}^{(1)} + 3 c_{\phi^4}^{(2)} + c_{\phi^4}^{(3)})\,-\frac{8}{3}g_2^2 \left[  c^{(1)}_{l^2 \phi^2 D^3} + c^{(2)}_{l^2 \phi^2 D^3} + 3( c^{(1)}_{q^2 \phi^2 D^3} + c^{(2)}_{q^2 \phi^2 D^3})  \right]_{\alpha_1,\alpha_1}\,,\\
 \dot{c}_{W^2\phi^2 D^2}^{(2)} &= -\frac{1}{24}g_2^2 (2 c_{\phi^4}^{(1)} + 3 c_{\phi^4}^{(2)} + c_{\phi^4}^{(3)})\,+\frac{2}{3}g_2^2 \left[c^{(1)}_{l^2 \phi^2 D^3} + c^{(2)}_{l^2 \phi^2 D^3} + 3( c^{(1)}_{q^2 \phi^2 D^3} + c^{(2)}_{q^2 \phi^2 D^3})  \right]_{\alpha_1,\alpha_1}\,,\\
 \dot{c}_{W^2\phi^2 D^2}^{(5)}&= \frac{2}{3}g_2^2 \left[c^{(3)}_{l^2 \phi^2 D^3} + c^{(4)}_{l^2 \phi^2 D^3} + 3( c^{(3)}_{q^2 \phi^2 D^3} + c^{(4)}_{q^2 \phi^2 D^3})  \right]_{\alpha_1,\alpha_1}\,,\\[0.3cm]
 \dot{c}_{WB\phi^2 D^2}^{(1)} &= -\frac{1}{12} g_1 g_2( c_{\phi^4}^{(2)} + c_{\phi^4}^{(3)}) -\frac{4}{3} g_1 g_2 \left[c^{(3)}_{l^2 \phi^2 D^3} + c^{(4)}_{l^2 \phi^2 D^3} -  c^{(3)}_{q^2 \phi^2 D^3} - c^{(4)}_{q^2 \phi^2 D^3}  \right]_{\alpha_1,\alpha_1}\,,\\
 \dot{c}_{WB\phi^2 D^2}^{(4)} & = \frac{1}{6} g_1 g_2( c_{\phi^4}^{(2)} + c_{\phi^4}^{(3)}) +\frac{8}{3} g_1 g_2 \left(c^{(3)}_{l^2 \phi^2 D^3} + c^{(4)}_{l^2 \phi^2 D^3} -  c^{(3)}_{q^2 \phi^2 D^3} - c^{(4)}_{q^2 \phi^2 D^3}  \right)_{\alpha_1,\alpha_1}\,,\\[0.3cm]
 \dot{c}_{G^2 \phi^2 D^2}^{(1)} & = -\frac{8}{3} g_3^2 \left(c^{(1)}_{u^2 \phi^2 D^3} + c^{(2)}_{u^2 \phi^2 D^3}+c^{(1)}_{d^2 \phi^2 D^3} + c^{(2)}_{d^2 \phi^2 D^3} + 2 c^{(1)}_{q^2 \phi^2 D^3} + 2 c^{(2)}_{q^2 \phi^2 D^3}  \right)_{\alpha_1,\alpha_1}   \,,\\
 \dot{c}_{G^2 \phi^2 D^2}^{(2)} &  =  \frac{2}{3} g_3^2  \left(c^{(1)}_{u^2 \phi^2 D^3} + c^{(2)}_{u^2 \phi^2 D^3}+c^{(1)}_{d^2 \phi^2 D^3} + c^{(2)}_{d^2 \phi^2 D^3} + 2 c^{(1)}_{q^2 \phi^2 D^3} + 2 c^{(2)}_{q^2 \phi^2 D^3}  \right)_{\alpha_1,\alpha_1} \,.
\end{align}

Essentially, all zeros in Tabs.~\ref{tab:dim8adm} and ~\ref{tab:dim8adm2} can be understood on the basis of the results in Ref.~\cite{Craig:2019wmo}. Within our field approach to renormalisation relying on the Green's basis of Ref.~\cite{Chala:2021cgt}, the only zero that is not manifest off-shell is the divergence of $\mathcal{O}_{W^2B\phi^2}^{(1)}$. Indeed, off-shell we have:
\begin{align}
 c_{W^2B\phi^2}^{(1)} &= \frac{g_2^2}{192\pi^2 \epsilon} c_{B\phi^4D^2}^{(1)} +\frac{g_2^2}{24 \pi^2 \epsilon} \left(c^{(1)}_{l^2 B \phi^2 D} + 3 c^{(1)}_{q^2 B \phi^2 D} \right)_{\alpha_1,\alpha_1}\nonumber\\
 & -\frac{g_1 g_2^2}{48 \pi^2 \epsilon} \left(c^{(3)}_{l^2 \phi^2 D^3} + 3 c^{(4)}_{l^2 \phi^2 D^3} - 5 c^{(3)}_{q^2 \phi^2 D^3} + c^{(4)}_{q^2 \phi^2 D^3} \right)_{\alpha_1,\alpha_1} \,, \\
 c_{WB\phi^2D^2}^{(11)} &= 0 \,,\\
 c_{WB\phi^2D^2}^{(13)} &= -\frac{g_2}{192\pi^2 \epsilon} c_{B\phi^4D^2}^{(1)}-\frac{g_2}{24 \pi^2 \epsilon} \left(c^{(1)}_{l^2 B \phi^2 D} + 3 c^{(1)}_{q^2 B \phi^2 D} \right)_{\alpha_1,\alpha_1}\nonumber\\
 & +\frac{g_1 g_2}{48 \pi^2 \epsilon} \left(c^{(3)}_{l^2 \phi^2 D^3} + 3 c^{(4)}_{l^2 \phi^2 D^3} - 5 c^{(3)}_{q^2 \phi^2 D^3} + c^{(4)}_{q^2 \phi^2 D^3} \right)_{\alpha_1,\alpha_1}\,. 
\end{align}
However, the Wilson coefficient  $c_{W^2B\phi^2}^{(1)}$  is shifted on-shell to
\begin{align}
 c_{W^2B\phi^2}^{(1)} \rightarrow c_{W^2B\phi^2}^{(1)} + \frac{g_2}{2}c_{WB\phi^2D^2}^{(11)} + g_2 c_{WB\phi^2D^2}^{(13)} \,,
\end{align}
thus making $c_{W^2B\phi^2}^{(1)}$ vanish on-shell.
It should be also emphasised that some anomalous dimensions arise only from redundant operators. For example, the physical interactions $\phi^4 D^4$ do not renormalise directly the operator $\mathcal{O}_{\phi^8}$,  because divergences of the latter are momentumless, whereas loops of the former involve always external momenta.

In Tabs.~\ref{tab:dim6adm} and ~\ref{tab:dim6adm2} we depict the same information as in Tabs.~\ref{tab:dim8adm} and ~\ref{tab:dim8adm2} but for the renormalisation of the bosonic SM Lagrangian terms and dimension-six interactions. 
In this case, the anomalous dimensions that differ substantially from naive power counting read:
\begin{align}
 \dot{c}_{\phi \square} &= 24 \lambda \mu^2 c_{\phi^4}^{(3)}+\frac{21}{2} g_2 \mu^2c_{W\phi^4D^2}^{(1)}+16\mu^2 c_{\phi^6}^{(2)} -16\mu^2 \left[ (c^{(1)}_{u^2 \phi^2 D^3}+c^{(2)}_{u^2 \phi^2 D^3})_{_{\alpha_1,\alpha_2}} \, y^u_{_{\alpha_3,\alpha_1}} (y^u)^*_{_{\alpha_2,\alpha_3}}\right.\nonumber\\
 & \left.+(c^{(1)}_{q^2 \phi^2 D^3}+c^{(2)}_{q^2 \phi^2 D^3}-c^{(3)}_{q^2 \phi^2 D^3}-c^{(4)}_{q^2 \phi^2 D^3})_{_{\alpha_1,\alpha_2}} \, y^u_{_{\alpha_2,\alpha_3}} (y^u)^*_{_{\alpha_3,\alpha_1}}\right]+ \cdots\,,\\[0.3cm]\nonumber
 \dot{c}_{\phi} &= 144 \lambda^2 \mu^2 c_{\phi^4}^{(3)}+14 g_1 \lambda \mu^2c_{B\phi^4D^2}^{(1)}+40 g_2 \lambda \mu^2 c_{W\phi^4D^2}^{(1)}\,-18 g_2^2 \mu^2 c_{W^2\phi^4}^{(1)}+52\lambda\mu^2 c_{\phi^6}^{(1)}+40\mu^2c^{8}_{\phi}\nonumber\\
 & -72 \lambda \mu^2 \left[ (c^{(3)}_{q^2 \phi^2 D^3}+c^{(4)}_{q^2 \phi^2 D^3})_{_{\alpha_1,\alpha_2}} \, y^u_{_{\alpha_2,\alpha_3}} (y^u)^*_{_{\alpha_3,\alpha_1}}  \right]- 12 \mu^2 (c^{(2)}_{q^2 \phi^4 D})_{_{\alpha_2,\alpha_1}} \, y^u_{_{\alpha_1,\alpha_3}} (y^u)^*_{_{\alpha_2,\alpha_3}} + \cdots \,,\\[0.cm]
 \dot{\lambda} &= -94 \lambda \mu^4 c_{\phi^4}^{(3)}-16 g_2 \mu^4c_{W\phi^4D^2}^{(1)} - 28 \mu^4 \left[ (c^{(3)}_{q^2 \phi^2 D^3}+c^{(4)}_{q^2 \phi^2 D^3})_{_{\alpha_1,\alpha_2}} \, y^u_{_{\alpha_2,\alpha_3}} (y^u)^*_{_{\alpha_3,\alpha_1}} \right]+ \cdots\,.
\end{align}
Likewise, the loop operators that are renormalised by (dimension-eight) tree level terms have the following RGEs:
\begin{align}
 \dot{c}_{\phi B} &= \mu^2\left[\frac{1}{4} g_1^2 \left(2 c_{\phi ^4}^{(1)}-3 c_{\phi ^4}^{(2)}+c_{\phi ^4}^{(3)}\right)-2g_1 c_{B\phi^4D^2}^{(1)}+12c_{B^2\phi^4}^{(1)}\right]\,,\\
 \dot{c}_{\phi \widetilde{B}} &= -2 \mu^2 \left(g_1 c_{B\phi^4D^2}^{(1)}-6c_{B^2\phi^4}^{(1)}\right)\,,\\
 \dot{c}_{\phi W} &= \mu^2\left[-\frac{g_2^2}{4} \left(c_{\phi ^4}^{(2)}-c_{\phi ^4}^{(3)} \right)-2g_2 c_{W\phi^4D^2}^{(1)}+12c_{W^2\phi^4}^{(1)}+4c_{W^2\phi^4}^{(3)}\right]\,,\\
 \dot{c}_{\phi \widetilde{W}} &= \mu^2\left[-2g_2 c_{W\phi^4D^2}^{(2)}+12c_{W^2\phi^4}^{(2)}+4c_{W^2\phi^4}^{(4)}\right]\,,\\
 \dot{c}_{\phi WB} &= \mu^2\left[\frac{g_1 g_2}{2} \left(c_{\phi ^4}^{(1)}-2c_{\phi ^4}^{(2)}+c_{\phi ^4}^{(3)}\right)-g_2 c_{B\phi^4D^2}^{(1)}-g_1 c_{W\phi^4D^2}^{(1)}+8c_{WB\phi^4}^{(1)}\right]\,,\\
 \dot{c}_{\phi \widetilde{W}B} &= -\mu^2\left[g_2 c_{B\phi^4D^2}^{(2)}+g_1 c_{W\phi^4D^2}^{(2)}-8c_{WB\phi^4}^{(2)}\right]\,,\\
 \dot{c}_{\phi G} &= 12\mu^2c_{G^2\phi^4}^{(1)}\,,\\
 \dot{c}_{\phi \widetilde{G}} &= 12\mu^2c_{G^2\phi^4}^{(2)}\,.
\end{align}
Notice that none of the dimension-eight fermionic interactions considered here renormalise lower-dimensional bosonic terms.

\section{Positivity bounds}
\label{sec:positivity}
Positivity bounds are restrictions on the form of the S-matrix derived from unitarity, analyticity and crossing. The best known example is the positivity of the forward scattering amplitude $\mathcal{A}(s,t=0)$ in $2\to 2$ processes, given by:
\begin{equation}\label{eq:positivity}
 \frac{d^2}{ds^2}\mathcal{A}(s,t=0)\bigg|_{s=0} \geq 0\,.
\end{equation}
If $\mathcal{A}(s,t=0)$ is analytical in $s=0$, then it admits an expansion in a neighbourhood of the origin reading $\mathcal{A}(s,t=0) = a_0+a_1s+a_2 s^2+...$. Hence, the equation above implies $a_2\geq0$.

The Wilson coefficients that enter in $a_2$ are those of dimension-eight (or pairs of dimension-six ones, etc.), from where it follows that certain combinations of dimension-eight terms are forced to be non-negative.
For example, by analysing the process $\phi\phi\to\phi\phi$, Ref.~\cite{Remmen:2019cyz} finds that:
\begin{align}~\label{eq:posh4}
c_{\phi^4 }^{(2)}\geq0\,,\\
c_{\phi^4 }^{(1)}+c_{\phi^4 }^{(2)}\geq0\,,\\
c_{\phi^4}^{(1)}+c_{\phi^4}^{(2)}+c_{\phi^4 }^{(3)}\geq0.
\end{align}
The problem arises 
 when the amplitude is not regular around $s=0$. This occurs, for example, when the leading contribution is due to loops involving massless particles, in which case there are branch cuts that extend all the way to the origin. In principle, the reasoning leading to Eq.~\eqref{eq:positivity} can be replicated upon giving a small mass $m$ to the massless states (hence regularising the singularity in the origin), which can be later taken to zero~\cite{Adams:2006sv}. However, this solution does not always imply that the conclusions derived in the absence of light loops remain valid in their presence.

In particular, for $\phi^4 D^4$ operators, it was shown recently  in Ref.~\cite{Chala:2021wpj} that the corresponding amplitude in the limit $m\to 0$ is actually dominated by the running of lower-dimensional operators ($\phi^4$ and $\phi^4 D^2$). It was explicitly shown that, in general, even if the $\phi^4 D^4$ operators fulfil the inequalities in Eqs.~\eqref{eq:posh4} at some heavy scale $\Lambda$ at which they are generated at tree-level, the positivity bounds are violated at scales $\tilde{\mu}<\Lambda$ upon evolving with the RGEs triggered by $\lambda$, $g_1$ and $g_2$. (Not so by gravity~\cite{Baratella:2021guc}.)

Positivity bounds exist also for the operators $X^2\phi^2 D^2$. They were obtained in Ref.~\cite{Bi:2019phv} upon inspection of the amplitude $V_1V_2\to V_1V_2$, with $V_i=W^\pm, Z, \gamma$, in the EW broken phase. The relevant operators in Ref.~\cite{Bi:2019phv} are dubbed $\mathcal{O}_{M,i}$, for $i=1,...,5,7$. The explicit form of those interactions (first derived in Ref.~\cite{Eboli:2006wa}) can be found also in that paper; here we simply specify how the corresponding Wilson coefficients (which are there called $f_{M,i}$) are related to the Wilson coefficients of the operators in our basis: 
\begin{align}\nonumber
 f_{M,0} &= -\frac{2}{g_2^2} c_{W^2\phi^2 D^2}^{(2)}\,,
 f_{M,1} = \frac{2}{g_2^2} (c_{W^2\phi^2 D^2}^{(1)}+c_{W^2\phi^2 D^2}^{(4)})\,,
 f_{M,2} = -\frac{4}{g_1^2} c_{B^2\phi^2 D^2}^{(2)}\,,\\
 f_{M,3} = \frac{4}{g_1^2} &c_{B^2\phi^2 D^2}^{(1)}\,,
 f_{M,4} = -\frac{4}{g_1 g_2} c_{WB\phi^2 D^2}^{(1)}\,,
 f_{M,5} = -\frac{8}{g_1g_2} c_{WB\phi^2 D^2}^{(4)}\,,
 f_{M,7} = \frac{4}{g_2^2} c_{W^2\phi^2 D^2}^{(4)}\,.
\end{align}
The equations (3.93) and (3.99) in Ref.~\cite{Bi:2019phv} constrain certain combinations of $f_{M,i}$. Translated to our basis, these relations read:
\begin{align}
 g_1^2c_{B^2\phi^2 D^2}^{(1)} + \mc{g_2^2}c_{W^2\phi^2 D^2}^{(1)} + 2\mc{g_1 g_2} c_{WB\phi^2 D^2}^{(4)} \leq 0\,,\label{eq:posI}\\
 \mc{g_1^2}c_{B^2\phi^2 D^2}^{(1)} + \mc{g_2^2}c_{W^2\phi^2 D^2}^{(1)} - 2 \mc{g_1g_2}c_{WB\phi^2 D^2}^{(4)} \leq 0\,,\\
 c_{W^2 \phi^2 D^2}^{(1)} \leq 0\,,\label{eq:bound4}\\
 g_1^2 c_{W^2 \phi^2 D^2}^{(1)} + 2 g_1 g_2 c_{WB \phi^2 D^2}^{(4)} + g_2^2 c_{B^2 \phi^2 D^2}^{(1)} \leq 0\,,\label{eq:bound5}\\
 g_1^2 c_{W^2 \phi^2 D^2}^{(1)} - 2 g_1 g_2 c_{WB \phi^2 D^2}^{(4)} + g_2^2 c_{B^2 \phi^2 D^2}^{(1)} \leq 0\label{eq:posF}\,.
\end{align}
One can also derive these constraints from the amplitudes $\phi V\to\phi V$ in the unbroken phase. For example, let us consider the process $\varphi_2 Z\to\varphi_2 Z$, with $\varphi_2$ being one of the  real degrees of freedom of the Higgs doublet, $\phi=\frac{1}{\sqrt{2}}(\varphi_1+i\varphi_2,\varphi_3+\varphi_4)^T$. We have:
\begin{align}
 -(g_1^2+g_2^2)\mathcal{A}(s) &= \left[g_2^2 c_{W^2\phi^2 D^2}^{(1)}+2g_1g_2 c_{WB\phi^2 D^2}^{(4)}+g_1^2 c_{B^2\phi^2 D^2}^{(1)}\right]\nonumber\\
 &\times \left[\epsilon^*_4 \cdot p_2\, \epsilon_2\cdot (p_1+p_3)\, s + \epsilon^*_4\cdot (p_1+p_3)\,\epsilon_2\cdot p_4\, s -2\epsilon^*_2\cdot\epsilon_4\, s^2\right]\,.
\end{align}
In the forward limit, $p_2\leftrightarrow p_4$, and so $\epsilon_4^*\cdot p_2$ and $\epsilon_2\cdot p_4$ vanish because the polarizations are transverse (the $Z$ is massless in the unbroken phase). Moreover, taking linearly polarized $Z$ bosons, we have $\epsilon_2=\epsilon_4=(0,a,b,0)$, and hence $\epsilon_2^*\cdot\epsilon_4=-(a^2+b^2)$. As a result, $\mathcal{A}$ depends only on $s$ despite involving particles with spin, as described in Ref.~\cite{Bellazzini:2016xrt}; more concretely:
\begin{align}
 \mathcal{A}(s) = -\frac{2 (a^2+b^2)}{g_1^2+g_2^2} \left[g_2^2 c_{W^2\phi^2 D^2}^{(1)}+2g_1g_2 c_{WB\phi^2 D^2}^{(4)}+g_1^2 c_{B^2\phi^2 D^2}^{(1)}\right] s^2\,,
\end{align}
which, following Eq.~\eqref{eq:positivity}, implies:
\begin{equation}
 g_2^2 c_{W^2\phi^2 D^2}^{(1)}+2g_1g_2 c_{WB\phi^2 D^2}^{(4)}+g_1^2 c_{B^2\phi^2 D^2}^{(1)} \leq 0\,.
\end{equation}
This is nothing but Eq.~\eqref{eq:posI}.

The benefit of working in the unbroken phase, though, is that analysing which lower-dimensional operators can dominate the amplitude $\phi V\to \phi V$ at low $s$ (namely proportional to $\log{s^2/\Lambda}$) is much easier. In particular,
and contrary to the $\phi^4 D^4$ instance, no relevant dimension-six operators are renormalised by tree-level interactions. Thus, to leading order in $g^2$, we have that for $s\to 0$:
\begin{equation}
 \mathcal{A}(s)\sim \dot{c}_{X^2\phi^2 D^2} \, \frac{s^2}{\Lambda^4}\, \log{\frac{s}{\Lambda^2}}\,.
\end{equation}
(The only relevant lower-dimensional couplings that get renormalised are the gauge couplings, but their contribution is proportional to the beta function of $g^2$, namely $\sim g^3$.) That is, because the only contribution is that of the running of $X^2\phi^2 D^2$, the RGEs of these operators must preserve positivity.

Let us ascertain this hypothesis by explicit calculation.

The operators $X^2\phi^2 D^2$ can not arise at tree-level in weakly-coupled UV completions of the SMEFT. So, their values at energies $\tilde{\mu}\ll\Lambda$ will be dominated by their running induced by tree-level operators, namely by $\phi^4 D^4$ \mc{and $\psi^2\phi^2 D^3$}. Using the RGEs in Eqs.~\eqref{eq:treeloppmixing} to leading-logarithm (note that 
the running of the SM gauge couplings can be ignored precisely because the $X^2 \phi^2 D^2$ operators vanish at tree-level at $\tilde{\mu}=\Lambda$), \mc{a sufficient condition for all inequalities above to hold is}:
\begin{align}\label{eq:posbound1}
 2 c_{\phi^4}^{(1)} + 3 c_{\phi^4}^{(2)} + c_{\phi^4}^{(3)}&\geq 0\,,\\
 c_{\phi^4}^{(1)} + 2 c_{\phi^4}^{(2)} + c_{\phi^4}^{(3)} &\geq 0\,,\\
 c_{\phi^4}^{(1)} + c_{\phi^4}^{(2)}  &\geq 0\,,\label{eq:posbound3}\\
 \left[\mc{c_{\psi_R^2\phi^2 D^3}^{(1)} + c_{\psi_R^2\phi^2 D^3}^{(2)}}\right]_{\alpha_1,\alpha_1} &\mc{\leq 0\,,}\label{eq:newposI}\\
 \left[\mc{c_{\psi_L^2\phi^2 D^3}^{(1)} + c_{\psi_L^2\phi^2 D^3}^{(2)} + c_{\psi_L^2\phi^2 D^3}^{(3)} + c_{\psi_L^2\phi^2 D^3}^{(4)}}\right]_{\alpha_1,\alpha_1}  &\mc{\leq 0\,,}\\
 \left[\mc{c_{\psi_L^2\phi^2 D^3}^{(1)} + c_{\psi_L^2\phi^2 D^3}^{(2)} - c_{\psi_L^2\phi^2 D^3}^{(3)} - c_{\psi_L^2\phi^2 D^3}^{(4)}}\right]_{\alpha_1,\alpha_1}  &\mc{\leq 0\,;}\label{eq:newposF} %
\end{align}
\mc{for $\psi_L=l,q$ and $\psi_R=e,u,d$. The Wilson coefficients above must be thought as evaluated in $\Lambda$.}
For example, Eq.~\eqref{eq:bound4} reads simply: 
\begin{align}
 c_{W^2\phi^2D^2}^{(1)}(\tilde{\mu}) &= c_{W^2\phi^2D^2}^{(1)}(\Lambda)-\frac{1}{16\pi^2}\dot{c}_{W^2\phi^2 D^2}^{(1)}(\Lambda)\log{\frac{\Lambda}{\tilde{\mu}}}<0\\
 &\Rightarrow \frac{1}{6} g_2^2 \bigg[2 c_{\phi^4}^{(1)} + 3 c_{\phi^4}^{(2)} + c_{\phi^4 D^4}^{(3)}\nonumber\\
 &\,\,\,\,\,\,\,\,\,\,\,\,\,\,\,\,\,-\frac{16}{3}\left(c_{l^2\phi^2D^3}^{(1)}+c_{l^2\phi^2D^3}^{(2)}+3c_{q^2\phi^2D^3}^{(1)}+3c_{q^2\phi^2D^3}^{(2)}\right)_{\alpha_1,\alpha_1}\bigg]\log{\frac{\Lambda}{\tilde{\mu}}} > 0\,,
\end{align}
and it is clear that the $g_2^2$ and the logarithm are positive. 
The relations in Eqs.~\eqref{eq:posbound1}--\eqref{eq:posbound3} are always fulfilled because, at tree-level, the four-Higgs operators satisfy the conditions in Eqs.~\eqref{eq:posh4}. The remaining inequalities, Eqs.~\eqref{eq:newposI}--\eqref{eq:newposF} are essentially equivalent to those quoted in Eq. (12) of Ref.~\cite{Li:2022tcz}. We have nevertheless checked their validity explicitly by studying the forward scattering amplitude for $\phi\psi\to\phi\psi$.
%
Thus, we have proven that the positivity constraints on the operators $X^2\phi^2 D^2$, unlike those for $\phi^4 D^4$~\cite{Chala:2021wpj}, remain valid at sufficiently small scales within one-loop accuracy. 

\section{Conclusions}
\label{sec:conclusions}
We have 
\mc{completed} the one-loop renormalisation of bosonic operators in the (lepton-number conserving) SMEFT to order $v^4/\Lambda^4$. This includes the running of operators triggered by pairs of dimension-six interactions, first computed in Ref.~\cite{Chala:2021pll}, as well as the renormalisation due to dimension-eight terms, which has been our focus for calculation within this work. We have relied heavily on the basis of Green's functions and the reduction of redundant operators onto physical ones derived in Ref.~\cite{Chala:2021cgt}. 
The current picture of one-loop renormalisation within the SMEFT is summarised in Tab.~\ref{tab:summary}.

Without entering into phenomenological considerations, there are several important consequences that can be derived from our results. To start with, we have found a number of anomalous dimensions that divert significantly from naive power counting. For example, $\dot{c}_{\phi^6}^{(1)} = -\frac{157}{16}g_2^4 c_{\phi^4 D^4}^{(3)} + 68\lambda c_{\phi^6}^{(1)}+\cdots$; the factors of $157/16$ and $68$ compensate partially the loop suppression. 
On a different note, there are tree-level dimension-eight interactions that mix into loop-level dimension-eight operators (this was first noticed in Ref.~\cite{Craig:2019wmo}) as well as into loop-level dimension-six terms \mc{(that we have unraveled here for the first time)}; %
an example of the latter is the renormalisation (proportional to $\mu^2$) of $X^2\phi^2$ interactions by $\phi^4 D^4$ operators. 
Finally, we have found the remarkable result that, unlike for $\phi^4 D^4$~\cite{Chala:2021wpj}, positivity bounds on $X^2\phi^2 D^2$ operators, first derived in Ref.~\cite{Bi:2019phv}, hold at all sufficiently small scales at one-loop accuracy. This strengthens the idea that the restrictions in Eqs.~\eqref{eq:posI}--\eqref{eq:posF} should be used as Bayesian priors in experimental fits aiming at measuring the values of the Wilson coefficients of quartic-gauge coupling operators, in line with Ref.~\cite{Zhang:2018shp}. 

Several future directions remain to be explored. First, one could compute the running driven by LNV and baryon-number violating (BNV) interactions. Although this effect is most probably negligible, due to the in principle huge scale of LNV and BNV, the possibility that these symmetries are broken at the TeV scale is not yet discarded; in which case the neutrino masses and the absence of proton decay would reflect strong GIM-like cancellations between different LNV and BNV operators~\cite{Dong:2011rh}.

One more step further requires renormalising the fermionic operators of the SMEFT. Our current work paves the way to this endeavour, since, in the field-theory approach to running, the divergences of fermionic interactions receive contributions from redundant bosonic operators, all of which we have computed here. As a third possible avenue, one could consider quantifying the impact of dimension-eight interactions, with and without quantum corrections, for constraining concrete models of new physics.

\begin{table}[t]
 \begin{center}
 \resizebox{1.\textwidth}{!}{\begin{tabular}{l|ccccccccccc}
   & $d_5$ & $d_5^2$ & $d_6$ & $d_5^3$ & $d_5\times d_6$ & $d_7$ & $d_5^4$ & $d_5^2\times d_6$ & $d_6^2$ & $d_5\times d_7$ & $d_8$\\
   \toprule
   $d_{\leq 4}$ (bosonic) &  &  & $\gtick$~\cite{Jenkins:2013zja} &  &   &  &  &  & $\gtick$~\cite{Chala:2021pll} &  & \colorbox{gray}{\textcolor{white}{This\, work}} \\
   $d_{\leq 4}$ (fermionic) &  &  & $\gtick$~\cite{Jenkins:2013zja} &  &   &  &  &  & $\rtick$ &  & $\rtick$ \\
   $d_5$ & $\gtick$~\cite{Chankowski:1993tx,Babu:1993qv,Antusch:2001ck} &  &  &  & $\gtick$~\cite{Chala:2021juk} & $\otick$~\cite{Chala:2021juk} &  &  &  & & \\
   $d_6$ (bosonic) &  & $\gtick$~\cite{Davidson:2018zuo} & $\gtick$~\cite{Jenkins:2013zja,Jenkins:2013wua,Alonso:2013hga} & & & &  & $\rtick$ & $\gtick$~\cite{Chala:2021pll} & $\rtick$ & \colorbox{gray}{\textcolor{white}{This\, work}} \\
   $d_6$ (fermionic) &  & $\gtick$~\cite{Davidson:2018zuo} & $\gtick$~\cite{Jenkins:2013zja,Jenkins:2013wua,Alonso:2013hga,Alonso:2014zka} & & & & & $\rtick$ & $\rtick$ & $\rtick$ & $\rtick$\\
   $d_7$ &  &  & & $\otick$~\cite{Chala:2021juk} & $\otick$~\cite{Chala:2021juk} & $\gtick$~\cite{Liao:2016hru,Liao:2019tep}\\
   $d_8$ (bosonic) &  &  & & & & & $\rtick$ & $\rtick$ & $\gtick$~\cite{Chala:2021pll} & $\rtick$ & \colorbox{gray}{\textcolor{white}{This\, work}}\\
   $d_8$ (fermionic) &  &  & & & & & $\rtick$ & $\rtick$ & $\rtick$ & $\rtick$ & $\otick$~\cite{AccettulliHuber:2021uoa}
   \\\bottomrule
 \end{tabular}}
 \caption{\it State of the art of the SMEFT renormalisation. The rows represent the operators (characterised by their dimension $d$) being renormalised, while the columns indicate the operators entering the loops. Blank entries vanish; a tick $\gtick$ represents that the complete contribution is known; the $\otick$ implies that only (but substantial) partial results have been already obtained; the $\rtick$ indicates that nothing, or very little, is known. The contribution made in this paper is marked by $\colorbox{gray}{\,\,\,}$.}\label{tab:summary}
 \end{center}
\end{table}

\section*{Acknowledgments}
We would like to thank José Santiago and Maria Ramos for  useful discussions.
This work has been partially funded by SRA under the grants 
PID2019-106087GB-C21/C22 (10.13039/501100011033), Junta de Andaluc\'ia under grants FQM 101, A-FQM-211-UGR18, P18-FR-4314 (FEDER) and A-FQM-467-UGR18 (FEDER) as well as by LIP (FCT, COMPETE2020-Portugal2020, FEDER, POCI-01-0145-FEDER-007334)  and by FCT under the project CERN/FIS-PAR/0032/2021 and under the grant SFRH/BD/144244/2019. 
ADC is also supported by the Spanish MINECO under the FPI programme.
MC is also supported by the Spanish MINECO under the Ram\'on y Cajal programme.

v3 : We thank Andreas Helset and Aneesh Manohar for pointing out numerical factor typos in some terms. These are corrected, and the Mathematica notebook in Github is revised. These results are cross-checked with Ref.~\cite{Assi:2023zid}, except for $\mathcal{O}_{l^2 H^4 D}^{(2)}$ and $\mathcal{O}_{q^2 H^4 D}^{(2)}$, as our definitions for these two operators differ from those in the referenced work. SDB acknowledges Mainz Institute of Theoretical Physics and EFT Foundations and Tools 2023 workshop organisers for hosting and hospitality where part of the communications are held.

{{\bf{Data Availability Statement:}} The datasets generated during the current study are available in the Github repository \href{https://github.com/SMEFT-Dimension8-RGEs}{https://github.com/SMEFT-Dimension8-RGEs} .}

\appendix
\section{Comparison with Ref.~\cite{AccettulliHuber:2021uoa}}\label{app:comparison}
The $\mathcal{O}(g^2,\lambda)$ contributions in the dimension-eight anomalous dimension matrix computed in this work (see Tab.~\ref{tab:dim8adm}) were previously calculated in Ref.~\cite{AccettulliHuber:2021uoa}, using a completely different approach based on on-shell amplitude methods. Here we check the consistency between the two results.

\mc{For simplicity, let us only focus} on the sub-matrix of the RGEs defined by the operators $\phi^8$, $\phi^6 D^2$, $\phi^4 D^4$, $X^2\phi^4$, $X\phi^4 D^2$. In the notation of Ref.~\cite{AccettulliHuber:2021uoa}, these operators are expanded by the minimal amplitudes $\mathcal{A}_{i}$, $i=1,19,18,44,45,46,11,10,8,6,9,7,5,4,3,2,43,41,42,40$, $25,24$. The rotation matrix that moves the corresponding Wilson coefficients in our basis to the Wilson coefficients in Ref.~\cite{AccettulliHuber:2021uoa} reads:
\begin{align}
 \mathcal{P} = \begin{pmatrix}
               1 & & & &\\
               & P_{\phi^6 D^2} & & &\\
               & & P_{\phi^4 D^4} & &\\
               & & & & P_{X^2\phi^4} &\\
               & & & & & P_{X\phi^4 D^2}
               \end{pmatrix}\,,
\end{align}
with
\begin{align}
 P_{\phi^6 D^2} = \begin{pmatrix}-1 & 2\\-1&1\end{pmatrix}\,,&\quad P_{\phi^4 D^2} = \begin{pmatrix}1 & 1 & 0\\ 1 & 0 & 1\\2 & 0 &0\end{pmatrix}\,,\\
 P_{X^2\phi^4} = \begin{pmatrix}Q & & & &\\& Q & Q & &\\ & & 4Q & &\\& & & Q&\\& & & & Q\end{pmatrix}\,, &\quad P_{X\phi^4 D^2} = \frac{1}{2}\begin{pmatrix}Q & Q &\\-Q & Q&\\& & 2Q\end{pmatrix}\,,
\end{align}
whereas 
\begin{equation}
 Q = \begin{pmatrix}1 & -i\\1 & i\end{pmatrix}\,.
\end{equation}
If, and only if, our computations are consistent with those in Ref.~\cite{AccettulliHuber:2021uoa}, our RGE matrix $\gamma$ truncated to order $\mathcal{O}(g^2,\lambda)$ is 
related to theirs, $\tilde{\gamma}$, through:
\begin{equation}\label{eq:xcheck}
 P^{-1}\tilde{\gamma}P = \gamma\,.
\end{equation}
One can indeed check, by direct calculation, that Eq.~\eqref{eq:xcheck} holds. 
\bibliographystyle{style} 
\bibliography{refs} 

\end{document}